\documentclass[prd,aps,twocolumn,floats]{revtex4}
\usepackage{latexsym}
\usepackage{amssymb}
\usepackage{amsmath}

\usepackage{epsfig}

\newcommand{\BI}{Bianchi identity }

\newcommand{\apJ}[3]{\apj {\bf #1}, #2 (#3) }

\newcommand{\cqg}[3]{Class. Quant. Grav. {\bf #1}, #2 (#3) }
\newcommand{\PRD}[3]{\prd{\bf #1}, #2 (#3) }

\newcommand{\grad}{\ensuremath{\vec{\nabla}}}
\newcommand{\lap}{\ensuremath{\vec{\nabla}^2}}
\newcommand{\TD}{\ensuremath{D}}
\newcommand{\oper}[1]{\ensuremath{\mathcal{#1}}}

\newcommand{\rhob}{\ensuremath{\bar{\rho}}}
\newcommand{\Pb}{\ensuremath{\bar{P}}}

\newcommand{\curv}[1]{\ensuremath{#1K}}

\newcommand{\PsiGI}{\ensuremath{\hat{\Psi}}}
\newcommand{\PhiGI}{\ensuremath{\hat{\Phi}}}

\begin{document}

\title{Consistent  cosmological modifications to the Einstein equations}

\author{Constantinos Skordis}
\affiliation{ Perimeter Institute, Waterloo, Ontario N2L 2Y5, Canada.} 
\email{cskordis@perimeterinstitute.ca}

\date{\today}

\renewcommand{\thefootnote}{\arabic{footnote}} \setcounter{footnote}{0}
%%%%%%%%%%%%%%%%%%%%%%%%%%%%%%%%%%%%%%%%%%%%%%%%%%%%%%%%%%%%%%%%%%%%%

%\abstract{
\begin{abstract}
General Relativity (GR) is a phenomenologically successful theory that rests on firm foundations, but  has not been tested on cosmological scales.
The advent of dark energy (and possibly even the requirement of cold dark matter), has increased the need for testing modifications to GR, as the inference of such otherwise
undetected fluids, depends crucially on the theory of gravity. In this work I outline 
a general scheme for constructing consistent and covariant modifications to the Einstein equations. This framework is such that there is a clear connection between the
modification and the underlying field content that produces it.
I conclude by a simple metric based modification of the fluctuation
equations for which the background is exact $\Lambda$CDM  and present its impact on observations of the cosmic microwave background radiation.
\end{abstract}

\maketitle
The theory of gravity plays a fundamental role in our modelling and understanding of the universe.
If we are to know the matter constituents of the universe,  we have to be sure we understand
what is the underlying gravitational theory.  Einstein's General Relativity (GR) has played a key role
in formulating modern cosmology, first as a smooth Friedmann-Lemaitre-Roberson-Walker (FLRW) spacetime, then
at the level of linearized fluctuations about this  spacetime. 

General relativity is a very solid "principle" theory from the theoretical point of
view, (and quite understandably the aesthetical point of view). The Lovelock-Grigore theorem~\cite{Lovelock,Grigore} asserts that
GR with a cosmological constant is unique under the following assumptions : geometry is Riemannian and the gravitational action depends only on the
metric, it is local and diffeomorphism invariant and leads to  2nd-order field equations.
 Relaxing any of these assumptions can lead to more general gravitational theories, e.g. adding extra fields~\cite{JFBD,TeVeS,AE},
having higher derivatives~\cite{Weyl}, having a pregeometry~\cite{pregeometry} or making the theory non-local\cite{non_local}.
This is not an exhaustive list but possible theories  fall into one or more categories above.

However as nice as we may think that GR is, the ultimate judge is experiment. Indeed, different aspects of GR have been vigorously tested
in the lab, in the solar system and with binary pulsars, all of which lie in the strong curvature regime (compared to cosmology).

 The  discovery that the expansion of the
universe is accelerating opens the possibility that general relativity breaks down on large scales
or low curvatures. It may also be that the apparent missing mass in the universe is not in the form of 
a cold dark matter particle but once again due to departures from general relativity.
This opens the need for cosmological tests of gravity, and much work has been carried out in this direction~\cite{PPF0,PPF1,PPF2} at various levels.
More recently Hu and Sawicki~\cite{HS}, and Hu~\cite{Hu}, have laid down a fully covariant formulation of 
modifications to gravity under well motivated assumtions.

In this work I outline a general scheme for constructing consistent modifications to the Einstein equations. 
The scheme is such that, one can clearly classify the modifications
according to whether they obey or violate diffeomorphism invariance,  need extra fields, or stem from higher derivative theories.
Indeed the advantage of this method is the direct connection between the field content and the modifications.
Additional assumptions as in~\cite{phipsi,HS,Hu} can always be used at the very end but we shall not consider this possibility here.
I close the paper by constructing the most general modification for which the FLRW background is exact $\Lambda$CDM and
illustrate the effects on observables in a simple subcase.

Following Hu~\cite{Hu} we start by putting the gravitational field equations in the form
\begin{equation}
G_{ab} = 8\pi G T_{ab}^{(known)}  + U_{ab}
\label{gen_U}
\end{equation}
Here $G_{ab}$ is the Einstein tensor for the universal matter metric $g_{ab}$, $T_{ab}^{(known)}$ is the stress-energy tensor of all \emph{known} forms
of matter (like baryons, photons and neutrinos),  $U_{ab}$ is a general tensor which encapsulates all the unknown fields/modifications
 and  can depend on $T_{ab}$ for each 
field and various combinations of metric functions (such as curvature tensors).
The assumption of upto 2nd order field equations, translates  to having only upto first derivatives of the extra field in $U_{0\nu}$
and upto 2nd derivatives in $U_{ij}$. Relaxing this
assumption is possible and will simply give higher order field equations, but one has to be cautious that quite generally higher derivative theories lead to
instabilites.

At this point we have to decide on the field content, i.e. whether $U_{ab}$ depends on additional fields or the metric alone.
In the former case, we must add one field at a time and ensure its energy conservation by applying the Bianchi identity 
which  directly translates to
\begin{equation}
 \nabla_a U^a_{\;\;b}   = 0
\end{equation}
and gives  a field equation for the extra field. 
 Violating the \BI leads to entirely arbitrary parametrizations and will not be considered.
The caveat of this approach is that in the case that more than two independent degrees of freedom are present in $U_{ab}$, one would have
to supply extra equations for these, not given by the \BI (see ~\cite{Hu2,Max,BFS} for examples).
 Finally, there could be interactions between  fields in $T_{ab}$ and $U_{ab}$. For simplicity  I do not consider this possibility further,
but it is straightforward to add.

We now split the dynamics of the problem in the background FLRW dynamics and their fluctuations about that background.
The FLRW metric is $ds^2 = a^2( -d\tau^2 + q_{ij} dx^i dx^j)$ where $\tau$ is the conformal time, $a$ is the scale factor and $q_{ij}$ is a
spacial metric of constant (dimensionless) curvature $K$.
The FLRW assumption  means that effectively we are considering a collection of scalar fields on the spacial hypersurface of
homogeneity and isotropy.  This boils down to requiring that the Lie derivative of the extra field vanishes for all six Killing vectors of the FLRW spacetime.
Examples are a scalar field $\phi(t)$, a vector field with components $A^\mu = (A(t), 0, 0, 0 )$ and a tensor field $X^a_{\;\;b}$ whose only non-vanishing 
components must be $X^0_{\;\;0} = -X(t)$ and $X^i_{\;\;j} = Y(t)\delta^i_{\;\;j}$. In the special case of a unit-timelike vector field $A(t)$ is pure gauge
and the contribution of such a field to the FLRW equations is generally given in terms of functions of  $a$, $\dot{a}$ and $\ddot{a}$~\cite{AE}.

Lets define $E_F= - a^2 G^0_{\;\;0}$ and $E_R$ such that $a^2 G^i_{\;\;j} = E_R \delta^i_{\;\;j}$, which are explicitely given by
$E_F =  3 \frac{\dot{a}^2}{a^2} +  3K $ and $ E_R =  - 2 \frac{\ddot{a}}{a} + \frac{\dot{a}^2}{a^2}  - K$.
The FLRW equations corresponding to  (\ref{gen_U}) are then simply written as  
$E_F = 8\pi G a^2 \sum_i \rho_i + a^2 X$
and
$E_R = 8\pi G a^2 \sum_i P_i + a^2 Y$.
  Applying the \BI then gives $\dot{E}_F + \frac{\dot{a}}{a}  (E_F +3E_R)  = 0$.
and   the fluid equation $\dot{\rho} + 3\frac{\dot{a}}{a}(\rho+P) = 0$.
Applying the Bianchi identity on $U_{ab}$  imposes
\begin{equation}
 \dot{X} + 3\frac{\dot{a}}{a}(X +  Y) =  0
\label{eq_bianchi_flrw}
\end{equation}
The above equation will give the background equation for the extra field, or additional constraints on $X$ and $Y$ in the absence thereof.
We see that any  FLRW  background can be modelled via one arbitrary function $Y(t)$ and a second function $X(t)$ found by solving  (\ref{eq_bianchi_flrw}).

Adding a scalar field amounts to letting $X = X(\phi,\dot{\phi},a,\dot{a})$ and $Y = Y(\phi,\dot{\phi},\ddot{\phi},a,\dot{a},\ddot{a})$, if we are to expect at most
2nd order field equations for the scalar. 
For example a canonical scalar field corresponds to $X =  \frac{1}{2a^2}\dot{\phi}^2 + V(\phi)$ and $Y = \frac{1}{2a^2}\dot{\phi}^2 - V(\phi) $
while the Jordan-Fierz-Brans-Dicke theory~\cite{JFBD} to $X = 8\pi G a^2 (e^{\phi} - 1)\rho + \frac{\omega}{2a^2} \dot{\phi}^2 + \frac{3}{a^2}\frac{\dot{a}}{a} \dot{\phi}   $ 
and $Y =  8\pi G a^2 (e^{\phi} - 1)P -\frac{1}{a^2}(\ddot{\phi} + \frac{\dot{a}}{a}\dot{\phi} - \frac{2+\omega}{2} \dot{\phi}^2)$.

Consider now the (scalar) fluctuations about the FLRW metric as
\begin{eqnarray}
ds^2 &=& - a^2(1 -  2 \Xi) dt^2 - 2a^2 (\grad_i \zeta) dt dx^i 
\nonumber
 \\ && + a^2[  (1 + \frac{1}{3}\chi ) q_{ij} + D_{ij} \nu ] dx^i dx^j
\end{eqnarray}
where $D_{ij} \equiv \grad_i \grad_j - \frac{1}{3} q_{ij} \grad^2$ is a spacial traceless derivative operator.
A perfect fluid is described at the fluctuation level by a density contrast $\delta$ ,
 momentum $\theta$ such that its total momentum is $u_i = a \grad_i \theta$, dimensionless pressure perturbation $\Pi$ such that $\delta T^i_{\;\;j} = \Pi\rho \delta^i_{\;\;j}$
 and shear $\Sigma$, such that the shear tensor is $\Sigma_{ij} = D_{ij} \Sigma$.

\begin{table}
\begin{tabular}{l|l}
$\Xi \rightarrow\Xi - \frac{\dot{\xi}}{a}$
& 
$\zeta \rightarrow \zeta + \frac{1}{a}\left[ \xi + \frac{\dot{a}}{a}\psi - \dot{\psi}\right]$
\\
$\chi \rightarrow \chi  + \frac{1}{a}\left[ 6 \frac{\dot{a}}{a}\xi + 2\lap \psi\right]$
&
$\nu \rightarrow \nu + \frac{2}{a}\psi$
\\
$V \rightarrow V + \frac{2}{a} \xi$
&
$J \rightarrow J + \frac{6}{a}  \frac{\dot{a}}{a}\xi$
\\
  $\theta \rightarrow \theta + \frac{1}{a}\xi$
&
$W \rightarrow W  + \frac{1}{a}\bigg[ 6 \frac{\dot{a}}{a}\dot{\xi} + 6 (\frac{\ddot{a}}{a} - 2\frac{\dot{a}^2}{a^2})\xi$
\\
$\delta \rightarrow \delta -\frac{3}{a}(1+w)\frac{\dot{a}}{a}\xi$
 & $ \qquad +2\lap \xi \bigg]$
\\
$E_\Theta \rightarrow E_\Theta + \frac{1}{a}(E_F + E_R)\xi$ 
& 
$\Pi \rightarrow \Pi + \frac{1}{a} \left[\dot{w} - 3w(1+w) \frac{\dot{a}}{a} \right]\xi$
\\
$E_P \rightarrow  E_P  + \frac{3}{a}\left[ \dot{E}_R -2\frac{\dot{a}}{a}  E_R \right] \xi$
&
$E_\Delta \rightarrow  E_\Delta - \frac{3}{a}\frac{\dot{a}}{a} (E_F + E_R)\xi$
\end{tabular}
\label{normal_gauge_trans}
\caption{Gauge transformations for the metric, fluid and Einstein tensor variables.}
\end{table}

We now decide whether the parametrization should obey diffeomorphism invariance. At the linearized level this is the requirement that all field equations
must be gauge form-invariant. Let us demonstrate what is gauge form-invariance for the case of standard GR coupled to a fluid.
Gauge transformations are infinitesimal diffeomorphisms generated by 
a vector field $\xi^a$ which can be  parametrized as $\xi_\mu = a ( -\xi, \grad_i \psi )$.
All perturbations apart from $\Sigma$ above are not gauge invariant but  transform with $\xi$ and $\psi$ as in table I.
Consider the $\delta G^0_{\;\;0}$ Einstein equation which is
\begin{eqnarray}
&& -\frac{1}{3} (\lap + \curv{3}) \left[ \chi -\lap \nu \right]
 + \frac{\dot{a}}{a}\left[ \dot{\chi} +2\lap\zeta \right] 
\nonumber \\&&
 \qquad \qquad 
+ 6 \frac{\dot{a}^2}{a^2} \Xi
= 8\pi G a^2 \rhob  \delta
\end{eqnarray}
If we perform a gauge transformation $X\rightarrow X'$ to all perturbations above, the $\delta G^0_{\;\;0}$ Einstein equation becomes
\begin{eqnarray}
&& -\frac{1}{3} (\lap + \curv{3}) \left[ \chi' -\lap \nu' \right]
 + \frac{\dot{a}}{a}\left[ \dot{\chi}' +2\lap\zeta' \right] 
\nonumber \\&&
 \qquad \qquad 
+ 6 \frac{\dot{a}^2}{a^2} \Xi'
= 8\pi G a^2 \rhob  \delta'
\end{eqnarray}
i.e. it retains its exact form (it is form-invariant), with the only change being a relabelling of the perturbations with $X\rightarrow X'$ for
each variable.
Gauge form-invariance always holds for all field equations which stem from a diffeomorphism invariant action,
no matter how complicated the theory is.

We can shortcut testing for gauge form-invariance as follows.  First define the three gauge non-invariant potentials $V\equiv \dot{\nu} +2 \zeta $, 
$J\equiv\chi - \lap \nu  $ and $W\equiv\dot{\chi} +2 \lap\zeta = \dot{J} + \lap V $, which are the only three combinations 
of metric variables appearing in the perturbed Einstein tensor. They transform only with the gauge variable $\xi$.
Then  define the two gauge-invariant potentials $\PhiGI \equiv -\frac{1}{6}J + \frac{1}{2}\frac{\dot{a}}{a} V$
and $\PsiGI \equiv - \Xi  - \frac{1}{2}\dot{V}  - \frac{1}{2}\frac{\dot{a}}{a} V$ .
We can now split the perturbed Einstein tensor into a gauge invariant and a gauge non-invariant part which involves the variable $V$. For
simplicity let us define $E_\Delta = -a^2\delta G^0_{\;\;0}$,   $E_\Theta$ is such that $-a^2\delta G^0_{\;\;i} = \grad_i E_\Theta$, $E_P = a^2 \delta G^i_{\;\;i}$
and $E_\Sigma$ is such that $ a^2\left[ \delta G^i_{\;\;j} - \frac{1}{3} \delta G^k_{\;\;k} \delta^i_{\;\;j}\right] = \TD^i_{\;\;j} E_\Sigma $.
Explicitely we get
\begin{eqnarray}
E_\Delta  &=& 2 (\lap + \curv{3}) \PhiGI
 - 6 \frac{\dot{a}}{a}(\dot{\PhiGI} +  \frac{\dot{a}}{a} \PsiGI)
\nonumber 
\\ &&
 -  \frac{3}{2}\frac{\dot{a}}{a} (E_F + E_R) V,
\label{E_Delta}
\end{eqnarray}
\begin{eqnarray}
E_\Theta &=& 2(\dot{\PhiGI} +  \frac{\dot{a}}{a}  \PsiGI) + \frac{1}{2}(E_F + E_R)V,
\label{E_Theta}
\end{eqnarray}
\begin{eqnarray}
E_P&=&
  6\frac{d}{dt}\left(\dot{\PhiGI} 
 + \frac{\dot{a}}{a} \PsiGI
\right)
+ 12\frac{\dot{a}}{a} (\dot{\PhiGI} + \frac{\dot{a}}{a} \PsiGI)
\nonumber \\ &&
 - 2(\lap  + \curv{3})(\PhiGI - \PsiGI)
 -3 (E_F+ E_R) \PsiGI 
\nonumber \\ &&
 + \frac{3}{2} [ \dot{E}_R -2 \frac{\dot{a}}{a}E_R ] V,
\label{E_P}
\end{eqnarray}
and
\begin{eqnarray}
E_\Sigma &=& \PhiGI - \PsiGI.
\label{E_Sigma}
\end{eqnarray}
The Einstein equations are then given as 
\begin{eqnarray}
E_\Delta &=& 8\pi G a^2 \rhob  \delta + U_\Delta
\label{E_Delta_eq}
\\
 E_\Theta &=&  8\pi G a^2 (\rhob + \Pb)\theta + U_\Theta
\label{E_Theta_eq}
\\
E_P &=& 24\pi G a^2 \rhob \Pi + U_P
\label{E_P_eq}
\\
E_\Sigma &=&   8\pi G (\rhob + \Pb)\Sigma + U_\Sigma
\label{E_Sigma_eq}
\end{eqnarray}
where the $U_i$ variables are defined in the same way as for the $E_i$ variables with $\delta G^a_{\;\;b}$ replaced by  $\delta U^a_{\;\;b}$.

The importance of gauge form-invariance cannot be overemphasized.  In fact  gauge form-invariance  severely constrains the terms involved.
For had we decided to let the RHS of the Einstein equations to be $ 8\pi G a^2 f_1(\tau) \delta$, $8\pi G a^2 f_2(\tau)\theta$  and
$ 24\pi G a^2 f_3(\tau) \Pi$ we would have found that by virtue of the background equations,  the Einstein equations are gauge form-invariant 
if and only if $f_1 = f_3 =  \rho$, and $f_2 = \rho+P$. 
A less trivial example is when $X = \frac{1}{r_c}\frac{\dot{a}}{a^2}$ for some scale $r_c$. If we decide that this term is due to metric modifications, 
then $U_\Delta = \frac{a}{r_c}(\frac{1}{6}\dot{J} + \frac{\dot{a}}{a}\Xi)$ and $U_\Theta =  -\frac{a^2}{3\dot{a}r_c}(\frac{1}{6}\dot{J} + \frac{\dot{a}}{a}\Xi)$.
If on the other hand we posit that it is due to a unit-timelike vector field perturbed as $A_i = a \grad_i \alpha$ then
a possibility is $U_\Delta = \frac{a}{r_c}(\frac{\ddot{a}}{a} - 2 \frac{\dot{a}^2}{a^2})\alpha$. In the last case
we pick higher derivatives, and a gauge-invariant counter-term must be added to cancel them~\cite{extra_skordis}.

Once we have added the gauge non-invariant terms and correctly fixed the functions multiplying them by requiring gauge form-invariance to hold,
we can proceed to add more gauge-invariant terms involving the extra fields and the gauge invariant potentials $\PhiGI$ and $\PsiGI$. 
If we want to consider only parametrizations which lead to 2nd order field equations in all variables, then we have to be
careful what terms we add and where. 
For example, when we add metric terms, we can add upto first derivatives in the two Einstein constraint
equations (\ref{E_Delta_eq}) and (\ref{E_Theta_eq}),  and up to second derivatives in the
two Einstein propagation equations (\ref{E_P_eq}) and (\ref{E_Sigma_eq}). Since $\PhiGI$  is of first order in the metric variables
while  $\PsiGI$ is of second order, we can add $\PhiGI$ in all four Einstein equations but $\dot{\PhiGI}$ and $\PsiGI$ 
only in the two propagation equations  (\ref{E_P_eq}) and (\ref{E_Sigma_eq}). Note that although $\dot{\PhiGI} + \frac{\dot{a}}{a} \PsiGI$ is of 1st-order in
the metric perturbations, 
it contains 2nd derivatives of the scale factor, and so it cannot be added to (\ref{E_Delta_eq}) and (\ref{E_Theta_eq}), unless
we relax the 2nd-order field equations constraint. The reason it is allowed in the  definition of $E_\Delta$ above is because, there are also 
2nd derivatives of the scale factor appearing in the gauge non-invariant part of $E_\Delta$ proportional to $V$. Thus when $E_\Delta$ is written in terms of the
actual metric potentials, the $\ddot{a}$ terms cancel and the final expression contains only first derivatives in all the variables. 
On the contrary, when we add
solely gauge-invariant terms we no longer have this luxury.

Let us also note that the gauge variable $\psi$ is not involved in the transformation of the Einstein tensor.
Thus if any extra field does transform with $\psi$ it will always appear in combination with $\nu$, $\zeta$ or $\chi$ in the
field equations, in a way that the whole combination does not transform with $\psi$.  An explicit example can be found in ~\cite{BFS}.

We finally  utilize the \BI which at the linearized level  simply translates in terms of the added  variables $U_\Delta$, $U_\Theta$, $U_P$ and
$U_\Sigma$ as
\begin{eqnarray}
   \dot{U}_\Delta +  \frac{\dot{a}}{a}  U_\Delta -  \lap U_\Theta +  \frac{1}{2} a^2 (X+Y) W +  \frac{\dot{a}}{a}  U_P &=&0
\qquad
\end{eqnarray}
and
\begin{eqnarray}
   \dot{U}_\Theta +2 \frac{\dot{a}}{a}  U_\Theta - \frac{1}{3}  U_P - \frac{2}{3} (\lap  + \curv{3} )U_\Sigma 
&&
\nonumber \\
 +   a^2 (X+Y) \Xi  &=& 0
\end{eqnarray}
This provides us with the field equations for the extra fields~\cite{extra_skordis}, or with additional constraints on the added functions in the absence thereof.

%\begin{figure}
%\epsfig{file=potentials.eps,height=3in}
%\caption{The time evolution of $g = \frac{\PhiGI - \PsiGI}{\PhiGI +\PsiGI}$  at $k=10^{-3}Mpc^{-1}$ for the simple modified gravity model in the text. The solid curve is the plain $\Lambda CMD$ model ($\beta = 0$), while the 
%dotted, dashed and dot-dashed curves are with $\beta = \{0.1, 0.5, 1\}$ respectively.  }
%\end{figure}

I now illustrate the above scheme, by finding the most general  diffeomorphism invariant 
modification to the Einstein equations for which the background cosmology is the plain
$\Lambda CDM$ model, no extra fields are present, and no higher derivative than two is present in the field equations.
Since there are no extra fields and the background is unchanged from $\Lambda CDM$  we can only add gauge invariant terms to Einstein equations
by setting $U_\Delta= \oper{A}\PhiGI$, $U_\Theta = \oper{B}\PhiGI$, $U_P = \oper{C}_1 \PhiGI + \oper{C}_2 \dot{\PhiGI} + \oper{C}_3  \PsiGI $ 
 and $ U_\Sigma = \oper{D}_1 \PhiGI + \oper{D}_2 \dot{\PhiGI} + \oper{D}_3  \PsiGI$, for operators $\oper{A}$, $\oper{B}$, $\oper{C}_i$ and $\oper{D}_i$.
Applying the \BI we get two equations involving $\dot{\PhiGI}$, $\PhiGI$ and $\PsiGI$ and consistency requires that
these equations must be satisfied whatever the values of  $\dot{\PhiGI}$, $\PhiGI$ and $\PsiGI$. 
 A sufficient condition is found by setting the coefficients of these terms to zero which gives $\oper{C}_3 = \oper{D}_3= 0$, $\oper{A} = -\frac{\dot{a}}{a}\oper{C}_2$
, $\oper{B} =  \frac{1}{3}\oper{C}_2 + \frac{2}{3} (\lap + \curv{3}) \oper{D}_2$
and the two differential equations $\dot{\oper{A}} + \frac{\dot{a}}{a}\oper{A}  - \lap \oper{B} + \frac{\dot{a}}{a}\oper{C}_1= 0$
and $\dot{\oper{B}} + 2 \frac{\dot{a}}{a} \oper{B} - \frac{1}{3}  \oper{C}_1 -  \frac{2}{3} (\lap + \curv{3})\oper{D}_1 = 0$.
A quick examination  reveals that if $\oper{A}$ and $\oper{B}$ are  both zero then we get exact GR.
The same holds for $\oper{D}_1$ and $\oper{D}_2$, hence a generic prediction of this kind of modification to GR is that $\PhiGI - \PsiGI$ should deviate 
from the GR value.
\begin{figure}
\epsfig{file=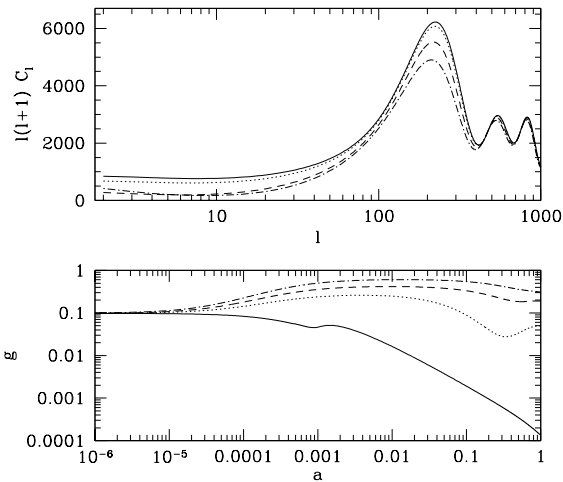,height=3.in}
\vspace{-1.27cm}
\caption{Upper panel: The CMB spectrum for the simple modified gravity model in the text. The solid curve is the plain $\Lambda CMD$ model ($\beta = 0$), while the 
dotted, dashed and dot-dashed curves are with $\beta = \{0.1, 0.5, 1\}$ respectively. Lower panel :
The time evolution of $g = \frac{\PhiGI - \PsiGI}{\PhiGI +\PsiGI}$ (as in ~\cite{HS,Hu}) at $k=10^{-3}Mpc^{-1}$  for the same set of models.
}
\end{figure}

Lets illustrate the effect on observables in a simple subcase for which $\oper{B} = \oper{C}_1 = \oper{D}_1 = 0$, 
 $\oper{A} = \frac{\beta H_0^2}{a}$, $\oper{C}_2 = -\frac{\beta H_0^2}{\dot{a}}$ 
and $\oper{D}_2 = \frac{\beta H_0^2}{2\dot{a}}\frac{1}{\lap}$. Thus,
we parametrize deviations from GR with a single dimensionless parameter $\beta$. We let $ \gamma = \frac{\beta H_0^2}{2k^2 a + \beta H_0^2}$
and find the Einstein equations in the synchronous gauge (defined as $\Xi = \zeta = 0$ and $\chi = h$, $-k^2\nu = h + 6\eta$). The $\dot{h}$ equation
is
\begin{equation}
\frac{\dot{a}}{a} \dot{h} = ( 1 - \gamma) 8\pi G a^2 \rho \delta + 2k^2 \eta - 6\gamma \frac{\dot{a}}{a} \dot{\eta}
\end{equation}
while the $\dot{\eta}$ equation is unchanged.
%and
%\begin{equation}
%2 \dot{\eta} =  8\pi G a^2 (\rho + P)\theta.
%\end{equation}

The perturbation equations were solved numerically  
 in both the synchronous and in the conformal Newtonian gauge for numerical consistency. The upper panel of Fig. 1 shows the CMB
angular power spectrum $l(l+1) C_l$ for a $\Lambda CDM$ universe ($\beta=0$) contrasted with non-zero $\beta$. We see that for this particular model,
the effect of non-zero $\beta$ is to decrease power on large scales including  even the first peak.
 The lower panel of Fig. 1 shows the time variation of $g = \frac{\PhiGI  - \PsiGI}{\PhiGI  + \PsiGI}$ 
for the same set of models at $k=10^{-3} Mpc^{-1}$. We see that like other modifications to gravity, the effect is to make $g$ grow.
In contrast to other parametrizations of modified gravity~\cite{PPF2,phipsi} however, the difference of $\PhiGI  - \PsiGI$ is  sourced by $\dot{\PhiGI}$ rather than
$\Phi$. 

This scheme can be used to consistently add terms to the Einstein equations which may or may not depend on extra fields, and possible metric functions.
The fluctuation equations can then be solved to obtain observable spectra. An important issue is to find predictions concerning these extra parameters
for upcoming precision experiments such as the Square Kilometer Array~\cite{SKA} and the EUCLID project.

{\it Acknoledgments: }
I'm greatful to M.Ba\~{n}ados, R. Durrer, P. Ferreira and W. Hu for important comments.
Research at Perimeter Institute for Theoretical Physics is supported in part by the Goverment of Canada through
NSERC and by the Province of Ontario through MRI.

\end{document}